\input harvmac
\input epsf

%
\let\includefigures=\iftrue
%
%
%
\newfam\black
\noblackbox
%
%
\includefigures
\message{If you do not have epsf.tex (to include figures),}
\message{change the option at the top of the tex file.}
\def\figin{\epsfcheck\figin}\def\figins{\epsfcheck\figins}
\def\epsfcheck{\ifx\epsfbox\UnDeFiNeD
\message{(NO epsf.tex, FIGURES WILL BE IGNORED)}
\gdef\figin##1{\vskip2in}\gdef\figins##1{\hskip.5in}
\else\message{(FIGURES WILL BE INCLUDED)}%
\gdef\figin##1{##1}\gdef\figins##1{##1}\fi}
\def\DefWarn#1{}
\def\N{{\cal N}}
\def\figinsert{\goodbreak\midinsert}
\def\ifig#1#2#3{\DefWarn#1\xdef#1{fig.~\the\figno}
\writedef{#1\leftbracket fig.\noexpand~\the\figno}%
\figinsert\figin{\centerline{#3}}\medskip\centerline{\vbox{\baselineskip12pt
\advance\hsize by -1truein\noindent\footnotefont{\bf
Fig.~\the\figno:} #2}}
\bigskip\endinsert\global\advance\figno by1}
\else
\def\ifig#1#2#3{\xdef#1{fig.~\the\figno}
\writedef{#1\leftbracket fig.\noexpand~\the\figno}%
\global\advance\figno by1} \fi

\def\tilde{\widetilde}

\def\yboxit#1#2{\vbox{\hrule height #1 \hbox{\vrule width #1
\vbox{#2}\vrule width #1 }\hrule height #1 }}
\def\fillbox#1{\hbox to #1{\vbox to #1{\vfil}\hfil}}
\def\ybox{{\lower 1.3pt \yboxit{0.4pt}{\fillbox{8pt}}\hskip-0.2pt}}

\def\rightarrowbox#1#2{
  \setbox1=\hbox{\kern#1{${ #2}$}\kern#1}
  \,\vbox{\offinterlineskip\hbox to\wd1{\hfil\copy1\hfil}
    \kern 3pt\hbox to\wd1{\rightarrowfill}}}

\def\vev#1{\langle{#1}\rangle}

\def\tilde{\widetilde}

\def\II{\relax{I\kern-.10em I}}

\def\bar{\overline}

\def\IZ{\relax\ifmmode\mathchoice
{\hbox{\cmss Z\kern-.4em Z}}{\hbox{\cmss Z\kern-.4em Z}}
{\lower.9pt\hbox{\cmsss Z\kern-.4em Z}} {\lower1.2pt\hbox{\cmsss
Z\kern-.4em Z}}\else{\cmss Z\kern-.4em Z}\fi}
\def\IB{\relax{\rm I\kern-.18em B}}
\def\IC{{\relax\hbox{$\inbar\kern-.3em{\rm C}$}}}
\def\ID{\relax{\rm I\kern-.18em D}}
\def\IE{\relax{\rm I\kern-.18em E}}
\def\IF{\relax{\rm I\kern-.18em F}}
\def\IG{\relax\hbox{$\inbar\kern-.3em{\rm G}$}}
\def\IGa{\relax\hbox{${\rm I}\kern-.18em\Gamma$}}
\def\IH{\relax{\rm I\kern-.18em H}}
\def\II{\relax{\rm I\kern-.18em I}}
\def\IK{\relax{\rm I\kern-.18em K}}
\def\IN{\relax{\rm I\kern-.18em N}}
\def\IP{\relax{\rm I\kern-.18em P}}

%
\def\inbar{\,\vrule height1.5ex width.4pt depth0pt}

\font\cmss=cmss10 \font\cmsss=cmss10 at 7pt
\def\IR{\relax{\rm I\kern-.18em R}}

\def\lp10{l_P^{10}}
\def\lp11{l_P^{11}}
\def\R11{R_{11}}

\newbox\tmpbox\setbox\tmpbox\hbox{\abstractfont
}
 \Title{\vbox{\baselineskip12pt\hbox to\wd\tmpbox{\hss
 hep-th/0501052} }}
 {\vbox{\centerline{Direct Proof Of Tree-Level Recursion Relation}
 \bigskip
 \centerline{In Yang-Mills Theory}
 }}
\smallskip
\centerline{Ruth Britto, Freddy Cachazo, Bo Feng, and Edward
Witten}
\smallskip
\bigskip
\centerline{\it School of Natural Sciences, Institute for Advanced
Study, Princeton NJ 08540 USA}
\bigskip
\vskip 1cm \noindent

\input amssym.tex

Recently, by using the known structure of one-loop scattering
amplitudes for gluons in Yang-Mills theory, a recursion relation
for tree-level scattering amplitudes has been deduced.  Here, we
give a short and direct proof of this recursion relation based on
properties of tree-level amplitudes only.

\Date{January 2005}
%

\lref\WittenNN{ E.~Witten, ``Perturbative Gauge Theory as a String
Theory in Twistor Space,'' hep-th/0312171.
}

\lref\ellissexton{R. K. Ellis and J. C. Sexton, "QCD Radiative
corrections to parton-parton scattering," Nucl. Phys.  {\bf B269}
(1986) 445.}

\lref\gravityloops{Z. Bern, L. Dixon, M. Perelstein, and J. S.
Rozowsky, ``Multi-Leg One-Loop Gravity Amplitudes from Gauge
Theory,"  hep-th/9811140.}

\lref\kunsztqcd{Z. Kunszt, A. Singer and Z. Tr\'{o}cs\'{a}nyi,
``One-loop Helicity Amplitudes For All $2\rightarrow2$ Processes
in QCD and ${\cal N}=1$ Supersymmetric Yang-Mills Theory,'' Nucl.
Phys.  {\bf B411} (1994) 397, hep-th/9305239.}

\lref\mahlona{G. Mahlon, ``One Loop Multi-photon Helicity
Amplitudes,'' Phys. Rev.  {\bf D49} (1994) 2197, hep-th/9311213.}

\lref\mahlonb{G. Mahlon, ``Multi-gluon Helicity Amplitudes
Involving a Quark Loop,''  Phys. Rev.  {\bf D49} (1994) 4438,
hep-th/9312276.}

\lref\klt{H. Kawai, D. C. Lewellen and S.-H. H. Tye, ``A Relation
Between Tree Amplitudes of Closed and Open Strings," Nucl. Phys.
{B269} (1986) 1.}

\lref\pppmgr{Z. Bern, D. C. Dunbar and T. Shimada, ``String Based
Methods In Perturbative Gravity," Phys. Lett.  {\bf B312} (1993)
277, hep-th/9307001.}

\lref\GiombiIX{ S.~Giombi, R.~Ricci, D.~Robles-Llana and
D.~Trancanelli, ``A Note on Twistor Gravity Amplitudes,''
hep-th/0405086.
}

\lref\WuFB{ J.~B.~Wu and C.~J.~Zhu, ``MHV Vertices and Scattering
Amplitudes in Gauge Theory,'' hep-th/0406085.
}

\lref\parke{S. Parke and T. Taylor, ``An Amplitude For $N$ Gluon
Scattering,'' Phys. Rev. Lett. {\bf 56} (1986) 2459; F. A. Berends
and W. T. Giele, ``Recursive Calculations For Processes With $N$
Gluons,'' Nucl. Phys. {\bf B306} (1988) 759. }

\lref\BrandhuberYW{ A.~Brandhuber, B.~Spence and G.~Travaglini,
``One-Loop Gauge Theory Amplitudes In N = 4 Super Yang-Mills From
MHV Vertices,'' hep-th/0407214.
}

\lref\CachazoZB{ F.~Cachazo, P.~Svr\v cek and E.~Witten, ``Twistor
space structure of one-loop amplitudes in gauge theory,''
hep-th/0406177.
}

\lref\RoiSpV{R.~Roiban, M.~Spradlin and A.~Volovich, ``A Googly
Amplitude From The B-Model In Twistor Space,'' JHEP {\bf 0404},
012 (2004) hep-th/0402016; R.~Roiban and A.~Volovich, ``All Googly
Amplitudes From The $B$-Model In Twistor Space,'' hep-th/0402121;
R.~Roiban, M.~Spradlin and A.~Volovich, ``On The Tree-Level
S-Matrix Of Yang-Mills Theory,'' Phys.\ Rev.\ D {\bf 70}, 026009
(2004) hep-th/0403190,
S.~Gukov, L.~Motl and A.~Neitzke,
``Equivalence of twistor prescriptions for super Yang-Mills,''
arXiv:hep-th/0404085,
I.~Bena, Z.~Bern and D.~A.~Kosower,
``Twistor-space recursive formulation of gauge theory amplitudes,''
arXiv:hep-th/0406133.
}

\lref\CachazoBY{ F.~Cachazo, P.~Svr\v cek and E.~Witten, ``Gauge
Theory Amplitudes In Twistor Space And Holomorphic Anomaly,''
hep-th/0409245.
}

\lref\GeorgiouBY{ G.~Georgiou, E.~W.~N.~Glover and V.~V.~Khoze,
``Non-MHV Tree Amplitudes In Gauge Theory,'' JHEP {\bf 0407}, 048
(2004), hep-th/0407027.
}

\lref\WuJX{ J.~B.~Wu and C.~J.~Zhu, ``MHV Vertices And Fermionic
Scattering Amplitudes In Gauge Theory With Quarks And Gluinos,''
hep-th/0406146.
}

\lref\WuFB{ J.~B.~Wu and C.~J.~Zhu, ``MHV Vertices And Scattering
Amplitudes In Gauge Theory,'' JHEP {\bf 0407}, 032 (2004),
hep-th/0406085.
}

\lref\GeorgiouWU{ G.~Georgiou and V.~V.~Khoze, ``Tree Amplitudes
In Gauge Theory As Scalar MHV Diagrams,'' JHEP {\bf 0405}, 070
(2004), hep-th/0404072.
}

\lref\CachazoDR{ F.~Cachazo, ``Holomorphic Anomaly Of Unitarity
Cuts And One-Loop Gauge Theory Amplitudes,'' hep-th/0410077.
}

\lref\seventree{F.~A. Berends, W.~T. Giele and H. Kuijf, ``Exact
And Approximate Expressions For Multi - Gluon Scattering,'' Nucl.
Phys. {\bf B333} (1990) 120.}

\lref\mangpxu{M. Mangano, S.~J. Parke and Z. Xu, ``Duality And
Multi - Gluon Scattering,'' Nucl. Phys. {\bf B298} (1988) 653.}

\lref\mangparke{M. Mangano and S.~J. Parke, ``Multiparton
Amplitudes In Gauge Theories,'' Phys. Rep. {\bf 200} (1991) 301.}

\lref\Bena{I. Bena, Z. Bern, D. A. Kosower and R. Roiban, ``Loops
in Twistor Space,'' hep-th/0410054.}

\lref\BrittoNJ{
R.~Britto, F.~Cachazo and B.~Feng,
``Computing one-loop amplitudes from the holomorphic anomaly of unitarity
cuts,''
arXiv:hep-th/0410179.
}

\lref\BidderTX{
S.~J.~Bidder, N.~E.~J.~Bjerrum-Bohr, L.~J.~Dixon and D.~C.~Dunbar,
``N = 1 supersymmetric one-loop amplitudes and the holomorphic anomaly of
unitarity cuts,''
arXiv:hep-th/0410296.
}


\lref\berends{F.~A.~Berends, R.~Kleiss, P.~De Causmaecker,
R.~Gastmans and T.~T.~Wu, ``Single Bremsstrahlung Processes In
Gauge Theories,'' Phys. Lett. {\bf B103} (1981) 124; P.~De
Causmaeker, R.~Gastmans, W.~Troost and T.~T.~Wu, ``Multiple
Bremsstrahlung In Gauge Theories At High-Energies. 1. General
Formalism For Quantum Electrodynamics,'' Nucl. Phys. {\bf B206}
(1982) 53; R.~Kleiss and W.~J.~Stirling, ``Spinor Techniques For
Calculating P Anti-P $\to$ W+- / Z0 + Jets,'' Nucl. Phys. {\bf
B262} (1985) 235; R.~Gastmans and T.~T. Wu, {\it The Ubiquitous
Photon: Heliclity Method For QED And QCD} Clarendon Press, 1990.}

\lref\xu{Z. Xu, D.-H. Zhang and L. Chang, ``Helicity Amplitudes
For Multiple Bremsstrahlung In Massless Nonabelian Theories,''
 Nucl. Phys. {\bf B291}
(1987) 392.}

\lref\gunion{J.~F. Gunion and Z. Kunszt, ``Improved Analytic
Techniques For Tree Graph Calculations And The G G Q Anti-Q Lepton
Anti-Lepton Subprocess,'' Phys. Lett. {\bf 161B} (1985) 333.}

\lref\BernKY{ Z.~Bern, V.~Del Duca, L.~J.~Dixon and D.~A.~Kosower,
``All Non-Maximally-Helicity-Violating One-Loop Seven-Gluon
Amplitudes In N = 4 Super-Yang-Mills Theory,'' hep-th/0410224.
}

\lref\BrittoNC{ R.~Britto, F.~Cachazo and B.~Feng, ``Generalized
Unitarity And One-Loop Amplitudes In N = 4 Super-Yang-Mills,''
hep-th/0412103.
}

\lref\RoibanIX{ R.~Roiban, M.~Spradlin and A.~Volovich,
``Dissolving N= 4 Loop Amplitudes Into QCD Tree Amplitudes,''
hep-th/0412265.
}

\lref\BrittoAP{ R.~Britto, F.~Cachazo and B.~Feng, ``New Recursion
Relations for Tree Amplitudes of Gluons,'' hep-th/0412308.
}

\lref\DixonWI{ L.~J.~Dixon, ``Calculating Scattering Amplitudes
Efficiently,'' hep-ph/9601359.
}

\lref\CachazoKJ{ F.~Cachazo, P.~Svr\v cek and E.~Witten, ``MHV
Vertices and Tree Amplitudes in Gauge Theory,'' hep-th/0403047.
}

\lref\BernBT{ Z.~Bern, L.~J.~Dixon and D.~A.~Kosower, ``All
Next-to-Maximally Helicity-Violating One-Loop Gluon Amplitudes in
N = 4 Super-Yang-Mills Theory,'' hep-th/0412210.
}


\newsec{Introduction}

Lately, there has been much renewed progress in understanding
tree-level and one-loop gluon scattering amplitudes in Yang-Mills
theory. Among other things,  a new set of recursion relations for
computing tree-level  amplitudes of gluons  has recently been
introduced \BrittoAP. These relations express any tree-level
amplitude of gluons as a sum over terms constructed from the
product of two subamplitudes with fewer gluons times a Feynman
propagator.   The subamplitudes are physical, on-shell amplitudes
with shifted momenta. These recursion relations were deduced by
using known properties of one-loop amplitudes to make inferences
about tree amplitudes.

A straightforward application of these recursion relations gives
new and unexpectedly simple forms for many amplitudes. Many of
these very compact forms have been obtained very recently
\refs{\BernKY,\BernBT,\RoibanIX} using somewhat related methods.

The recursion relations can be schematically written as follows
\eqn\qeqi{ A_n = \sum_r A^h_{r+1}{1\over P_r^2} A^{-h}_{n-r+1}. }
Here, for any positive integer $s$, $A_s$ denotes the tree-level
scattering amplitudes for $s$ cyclically ordered gluons.  In
writing a recursion relation for $A_n$, one ``marks'' two of the
gluons and sums over products of subamplitudes, with $r$ external
gluons on one side, $n-r$ external gluons on the other side, and
one internal gluon connecting them, and with the two marked or
reference gluons being on opposite sides. (The sum in \qeqi\ is
really a sum over decompositions with one marked gluon on each
side, not just a sum over $r$.) $P$ is the momentum and $h$ the
helicity of the internal gluon. Momenta are shifted so that this
gluon as well as the external ones are on-shell.

\ifig\kkmbp{Pictorial representation of the recursion relation
\qeqi. Thick lines represent the two marked gluons. The choice of
marked gluons here is in anticipation of the discussion in section
2. The sum is over all cyclically ordered distributions of gluons
with at least two gluons on each subamplitude and over the two
choices of helicity for the internal gluon.}
{\epsfxsize=0.85\hsize\epsfbox{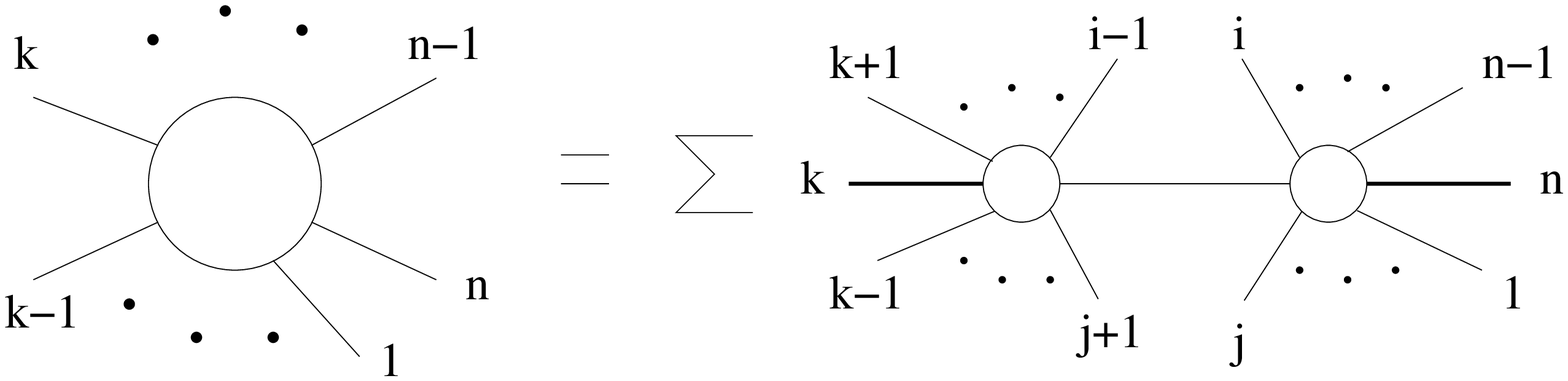}}

In \BrittoAP, an outline of a proof of this formula was given by
using a recently discovered method for computing one-loop
amplitude in $\N=4$ gauge theory \BrittoNC\ combined with the IR
behavior of the amplitudes.

However, the simplicity of the recursion relation \qeqi\ begs for
a  more direct and transparent derivation. In particular, one
suspects that there should be a derivation that uses properties of
tree amplitudes only, rather than deducing properties of tree
amplitudes from properties of loop amplitudes.

The aim of this note is to provide such a proof. The proof only
uses basic facts about tree diagrams, such as the fact that their
singularities come only from poles of internal propagators, plus
the description of tree amplitudes via MHV diagrams \CachazoKJ,
which we use at one step in the proof to show that the tree
amplitudes vanish in a certain limit. (For one arrangement of
helicities, we also prove this directly from standard Feynman
diagrams.) The recursion relations of \BrittoAP\ use two adjacent
gluons of opposite helicity as reference gluons. Our proof shows
that the recursion relations can also be defined for reference
gluons of the same helicity, and that the gluons do not have to be
adjacent.

We conclude this paper by using the BCF recursion relations to show
that MHV tree diagrams give the same Yang-Mills tree amplitudes as Feynman
diagrams.

\newsec{Derivation of the BCF Recursion Relations}

Gluon tree amplitudes are most conveniently written using the
spinor-helicity formalism \refs{\berends, \xu, \gunion}. In a
nutshell, the idea is that in four dimensions any null vector $p$
can be written as a bispinor, $p_{a\dot a} =
\lambda_a\tilde\lambda_a$. The inner product of vectors can be
written in terms of the natural inner product of spinors
$\vev{\lambda, ~\lambda'} = \epsilon_{ab}\lambda^a\lambda'^b$ and
$[\tilde\lambda, ~\tilde\lambda'] = \epsilon_{\dot a\dot
b}\tilde\lambda^{\dot a}\tilde\lambda^{\dot b}$. More explicitly,
if $q_{a\dot a}=\lambda'_a\tilde\lambda'_{\dot a}$, then $2p\cdot
q = \vev{\lambda, ~\lambda'}[\tilde\lambda, ~\tilde\lambda']$. It
turns out that polarization vectors also have a representation in
terms of spinors and the full amplitude becomes a rational
function of spinor products (for a review see \DixonWI).

Consider a tree-level amplitude $A(1,2,\ldots ,n-1,n)$ of $n$
cyclically ordered gluons, with any specified helicities. Denote
the momentum of the $i^{th}$ gluon by $p_i$ and the corresponding
spinors by $\lambda_i$ and $\tilde\lambda_i$.  Thus, $p_i^{a\dot
a} =\lambda_i^a\tilde\lambda_i^{\dot a}$.

In what follows, we single out two of the gluons for special
treatment.  Using the cyclic symmetry, without any loss of
generality, we can take these to be gluons $k$ and $n$. Introduce
a complex variable $z$, and let
\eqn\deftwo{ \eqalign{ p_{k}(z) & = \lambda_{k}(\tilde\lambda_{k}
- z\tilde\lambda_n) \cr p_n(z) & = (\lambda_n +
z\lambda_{k})\tilde\lambda_n } } We leave the momenta of the other
gluons unchanged, so $p_s(z)=p_s $ for $s\not= k,n$.
 In effect, we have made the
transformation
\eqn\thetrans{\tilde\lambda_k\to\tilde\lambda_k-z\tilde\lambda_n,~~\lambda_n\to\lambda_n+z\lambda_k,}
with $\lambda_k$ and $\tilde\lambda_n$ fixed. Note that $p_k(z)$
and $p_n(z)$ are on-shell for all $z$, and $p_k(z)+p_n(z)$ is
independent of $z$. As a result, we can define the following
function of a complex variable $z$,
\eqn\defz{A(z) = A(p_1,\ldots ,p_{k-1}, p_{k}(z), p_{k+1},\ldots
,p_{n-1}, p_n(z)). }
The right hand side is a physical, on-shell amplitude for all $z$.
Momentum is conserved and all momenta are on-shell.

For any $z\not=0$, the deformation \deftwo\ does not make sense
for real momenta in Minkowski space, as it does not respect the
Minkowski space reality condition $\tilde\lambda = \pm
\bar\lambda$. However,
\deftwo\ makes perfect sense for complex momenta or  (if $z$ is real) for
real momenta in signature $++-\,-$.   In any case, we think of
$A(z)$ as an auxiliary function. In the end, all answers are given
in terms of spinor inner products and are valid for any signature.
In the derivation of recursion relations, it will be necessary to
assume that the helicities $(h_k,h_n)$ are $(-,+)$, $(+,+)$, or
$(-,-)$.  To get a recursion relation in the remaining case
$(+,-)$, we use the cyclic symmetry to exchange the roles of $k$
and $n$, or equivalently, we exchange the roles of $\lambda$ and
$\tilde\lambda$ in \deftwo.

$A(z)$ is a rational function of $z$. To see this, note that the
original tree-level amplitude is a rational function of spinor
products, as we recalled above. Since the $z$ dependence only
enters via the shift $\tilde\lambda_k \to \tilde\lambda_k -
z\tilde\lambda_n$ and $\lambda_n \to \lambda_n + z \lambda_k$,
$A(z)$ is clearly rational in $z$.

In fact, more specifically, for generic external momenta, $A(z)$
has only simple poles as a function of $z$. Singularities come
only from the poles of a propagator in a Feynman diagram. As we
will see, each propagator gives only a single simple pole, and for
generic external momenta, distinct propagators give poles at
distinct values of $z$.

To explain these statements, recall first that the momentum
flowing through a propagator in a tree diagram is always a sum of
external momenta. In Yang-Mills theory, tree diagrams are planar,
and the momentum in a propagator is always a sum of momenta of
{\it adjacent} external particles, say $P_{ij}=p_i+\dots+p_j$ for
some $i,j$ with $j>i$. A propagator with this momentum is
$1/P_{ij}^2$. At nonzero $z$, this becomes $1/P_{ij}(z)^2$ with
$P_{ij}(z)=p_i(z)+\dots + p_j(z)$.  In our problem, as $p_s$ is
independent of $z$ for $s\not= k,n$, and $p_k(z)+p_n(z)$ is
independent of $z$, $P_{ij}(z)$ is completely independent of $z$
if both $k$ and $n$ or neither of them are in the range from $i$
to $j$.  We consider the remaining cases that one of $k,n$ is in
this range and the other is not.  By momentum conservation, we
could replace $p_i+\dots +p_j$ by $-(p_{j+1}+\dots +p_{i-1})$. So
there is no essential loss of generality in assuming that $n$ is
in the range from $i$ to $j$ while $k$ is not.  In this case,
$P_{ij}(z)=P_{ij}+z\lambda_k\tilde\lambda_n$, so $P_{ij}(z)^2
=P_{ij}^2 - z \langle\lambda_k|P_{ij}|\tilde \lambda_n]$ (where
for any spinors $\lambda,\tilde\lambda$ and vector $p$, we define
$\langle \lambda|p|\tilde\lambda]=- p_{a\dot
a}\lambda^a\tilde\lambda^{\dot a}$).  Clearly then, the propagator
$1/P_{ij}(z)^2$ has only a single, simple pole, which is located
at $z_{ij}= P_{ij}^2/\langle\lambda_k|P_{ij}|\tilde\lambda_n]$.
For generic external momenta, the $z_{ij}$ for distinct pairs
$i,j$ are distinct. These poles are the only singularities of
$A(z)$. So $A(z)$, as claimed, has only simple
poles.\foot{Incidentally, in this argument, there is no need to
distinguish collinear singularities (the cases $j=i+1$ and
$j=i-3$) from multiparticle singularities (the other cases). For
complex momenta, these can all be treated alike.}

In section 3, we use MHV tree diagrams to prove that $A(z)$
vanishes for $z\to\infty$ as long as the helicities of particles
$k$ and $n$ are $(-,+)$, $(+,+)$, or $(-,-)$.  (As explained
above, in the remaining case, one should make a slight
modification of \deftwo.)  A rational function $A(z)$ that
vanishes at infinity and whose only singularities are simple poles
at $z=z_{ij}$ has an expansion
\eqn\collo{A(z)=\sum_{i,j}{c_{ij}\over z-z_{ij}},} where  $c_{ij}$
are the  residues of the poles. From the above discussion, the sum
over $i$ and $j$ runs over all pairs such that $n$ is in the range
from $i $ to $j$ while $k$ is not.

The physical scattering amplitude that we want to calculate is
simply $A=A(0)$.  In terms of the poles and residues, it is
\eqn\bollo{A=-\sum_{i,j}{c_{ij}\over z_{ij}}.} This is obtained
from \collo\ by setting $z$ to zero in the denominators without
changing the numerators.  As we will now see, this formula is
equivalent to the BCF recursion relation.

In fact, it is easy to describe the residue of the pole at
$z=z_{ij}$.  To get a pole at $P_{ij}^2(z)=0$, a tree diagram must
contain a propagator that divides it into a ``left,'' containing
all external gluons not in the range from $i$ to $j$, and a
``right,'' containing all external gluons that are in that range.
See figure 1. The internal line connecting the two parts of the
diagram has momentum $P_{ij}(z)$, and we need to sum over the
helicity $h=\pm$ at, say, the left of this line. (The helicity at
the other end is opposite.)  The contribution of such a diagram is
$\sum_h A_L^h(z)A_R^{-h}(z)/P_{ij}(z)^2$, where $A_L^h(z)$ and
$A_R^{-h}(z)$ are the amplitudes on the left and the right with
indicated helicities. Since the denominator $P_{ij}(z)^2$ is
linear in $z$, to obtain the function $c_{ij}/(z-z_{ij})$ that
appears in \collo, we simply must set $z$ equal to $z_{ij}$ in the
numerator. When we do this, the internal line becomes on-shell,
and the numerator becomes a product
$A_L^h(z_{ij})A_R^{-h}(z_{ij})$ of physical, on-shell scattering
amplitudes.

The formula \collo\ for the function $A(z)$ therefore becomes
\eqn\gollo{A(z)=\sum_{i,j}\sum_h{A_L^h(z_{ij})A_R^{-h}(z_{ij})\over
P_{ij}(z)^2}.} To get the physical scattering amplitude \bollo, we
just need to set $z$ to zero in the denominator without touching
the numerator.  Hence,
\eqn\wollo{A=\sum_{i,j}\sum_h{A_L^h(z_{ij})A_R^{-h}(z_{ij})\over
P_{ij}^2}.} This is the BCF recursion relation.

\newsec{Vanishing At Infinity and MHV Diagrams}

In this section, we complete the proof by showing that $A(z)$
vanishes as $z\to \infty$ if $(h_{k},h_n)$ is equal to $(-,+)$,
$(+,+)$ or $(-,-)$, or more simply if $h_n=+$ or $h_k=-$.

The proof uses the MHV diagram construction of Yang-Mills tree
amplitudes  \CachazoKJ.   In this construction, one computes tree
amplitudes from tree-level Feynman diagrams in which the vertices
are MHV amplitudes, continued off-shell in a suitable fashion, and
the propagators are ordinary Feynman propagators.  We will present
the argument assuming that $h_n=+$, in which case, we can make the
argument using ordinary (``mostly plus'') MHV vertices. For
$h_k=-$, one makes the same argument using Feynman diagrams with
opposite helicity MHV vertices.

As a warmup, let us suppose that the $n$-gluon amplitude of
interest is actually an MHV amplitude.  Then
\eqn\hubu{A(z)={\langle\lambda_r,\lambda_s\rangle^4\over\prod_{i=1}^n
\langle\lambda_i,\lambda_{i+1}\rangle},} where $\lambda_n$ depends
on $z$ as in \thetrans, while the other $\lambda$'s are
independent of $z$.
 In \hubu, $r,s$ are the two
gluons of negative helicity; the others all have positive
helicity.  As long as gluon $n$ has $+$ helicity, $\lambda_n$ does
not appear in the numerator of $A(z)$, which therefore is
independent of $z$.  The denominator, on the other hand, is a
non-trivial polynomial in $z$, because of the factors $\langle
\lambda_{n-1},\lambda_n\rangle$ and
$\langle\lambda_n,\lambda_1\rangle$, at least one of which
(depending on $k$) has a nontrivial dependence on $z$.   Hence, if
$h_n=+$, $A_n(z)\to 0$ for $z\to\infty$.  This argument would
clearly fail if gluon $n$ had negative helicity. (For an opposite
helicity MHV amplitude, a similar argument shows that $A(z)$
vanishes for large $z$ if gluon $k$ has negative helicity.)

Now consider a general MHV tree diagram.  Its contribution to the
scattering amplitude is a product of off-shell MHV tree
amplitudes, times Feynman propagators $1/P^2$.  The propagators
are independent of $z$ or vanish for $z\to \infty$; in fact, their
behavior was analyzed in section 2.  It will suffice, therefore,
to show that the product of off-shell MHV tree amplitudes vanishes
for $z\to\infty$.  The key point is to show that if we set up the
MHV tree diagrams properly, the off-shell continuation used to
define the vertices does not spoil the behavior found in the last
paragraph.

A lightlike vector $p$ has a factorization $p_{a\dot
a}=\lambda_a\tilde\lambda_{\dot a}$, but there is no such
factorization for a vector $P$ that is not lightlike.  In defining
MHV tree diagrams, one needs to define a positive helicity spinor
$\lambda$ associated with each internal momentum $P$ in a Feynman
graph.  To do this one introduces an arbitrary and fixed negative
chirality spinor $\eta$ and defines $\lambda_a=P_{a\dot
a}\eta^{\dot a}$.   The individual MHV tree diagrams give
amplitudes that depend on $\eta$, but the sum does not \CachazoKJ.

In our present problem, it is extremely convenient to pick
$\eta=\tilde\lambda_n$.  The point is that this causes $\lambda$
to be independent of $z$ for each internal gluon, a fact that we
can show as follows. Each internal momentum $P$ is a sum
$p_i+p_{i+1}+\dots +p_j$, for some $i$ and $j$.  As in section 2,
the $z$-dependence, if any, of such a sum is proportional to
$\lambda_k\tilde\lambda_n$, so if $\eta=\tilde\lambda_n$, then
$\lambda_a=P_{a\dot a}\eta^{\dot a}$ is independent of $z$ for all
internal gluons. In this respect, the internal gluons are no
different from the external gluons other than gluon $n$.  Hence
(with our choice of $\eta$), in an off-shell MHV amplitude that
appears as a vertex in an MHV tree diagram, all $\lambda$'s except
$\lambda_n$, whether associated with internal or external gluons,
are independent of $z$.   With this at hand, the same analysis we
used for physical MHV tree amplitudes shows that, with gluon $n$
assumed to have positive helicity, the off-shell MHV vertex
containing gluon $n$ vanishes for $z\to\infty$. The other MHV
vertices are clearly independent of $z$.  So the product of the
MHV vertices vanishes for $z\to\infty$, and our argument is
complete.

\subsec{Analysis Using Standard Feynman Diagrams}

This argument, though brief, raises the question of whether MHV
tree diagrams are essential or the same result can be deduced from
more standard methods. Here we will give an alternative argument
based on ordinary Feynman diagrams for the case that the
helicities are $(h_k,h_n)=(-,+)$.

Recall that any Feynman diagram contributing to the amplitude
$A(z)$ is linear in the polarization vectors $\epsilon_{a\dot a}$
of the external gluons. Polarization vectors of gluons of negative
and positive helicity and momentum $p_{a\dot a}= \lambda_a
\tilde\lambda_{\dot a}$ can be written respectively as follows,
\eqn\poli{\epsilon_{a\dot a}^{(-)} = {\lambda_a\tilde\mu_{\dot
a}\over [\tilde\lambda , \tilde\mu ] }, \qquad \epsilon_{a\dot
a}^{(+)} = {\mu_a \tilde\lambda_{\dot a}\over \vev{\mu,
\lambda}},}
where $\mu$ and $\tilde\mu$ are fixed reference spinors.

Only the polarization vectors of gluons $k$ and $n$ can depend on
$z$. Consider the $k^{th}$ gluon first. Recall that $\lambda_k$
does not depend on $z$ and $\tilde\lambda_k(z)$ is linear in $z$.
Since $h_k=-$, it follows from \poli\ that $\epsilon^{(-)}_k$ goes
as $1/z$ as $z\to \infty$. A similar argument leads to
$\epsilon^{(+)}_n \sim 1/z$.

The remaining pieces in a Feynman graphs are the propagators and
vertices. It is clear that the vanishing of $A(z)$ as $z\to
\infty$ can only be spoiled by the momenta from the cubic
vertices, since the quartic vertices have no momentum factors and
the propagators can only vanish for $z\to\infty$.

Let us now construct the most dangerous class of graphs and show
that they vanish precisely as $1/z$.

The $z$ dependence in a tree diagram ``flows" from the $k^{th}$
gluon to the $n^{th}$ gluon in a unique path of propagators. Each
such propagator contributes a factor of $1/z$.  If there are $r$
such propagators, the number of cubic vertices through which the
$z$-dependent momentum flows is at most $r+1$. (If all vertices
are cubic, then
 starting from the $k^{th}$ gluon, we
find a cubic vertex and then a propagator, and so on. The final
cubic vertex is then joined to the $n^{th}$ gluon.) So the
vertices and propagators give a factor that grows for large $z$ at
most linearly in $z$.

As the product of polarization vectors vanishes as $1/z^2$, it
follows  that for this helicity configuration, $A(z)$ vanishes as
$1/z$ for $z\to \infty$.

\newsec{Proof Of MHV Recursion Relations}

For the $(+,-)$ helicity pair, we have shown that the generalized 
amplitude $A(z)$ vanishes at infinity,
and hence the BCF recursion relations are obeyed, using either standard
Feynman diagrams or the MHV tree diagrams.

It follows that the BCF relations for $(+,-)$ helicity are satisfied for 
either the amplitudes
computed using Feynman diagrams or the amplitudes computed using
MHV tree diagrams.
The $(+,-)$ relation is enough to determine the amplitude recursively, 
given
that the all $+$ and all $ -$ amplitudes vanish. Hence, the MHV tree 
diagrams give the same amplitudes
as the Feynman diagrams.

In making this argument for the MHV tree diagrams, one needs to know that
MHV tree diagrams have only the physical singularities of the Feynman 
diagrams.
That the physical singularities appear correctly was shown in section 4 
of \CachazoKJ.   Individual MHV tree diagrams
have additional unphysical singularities of the form $1/\langle 
\lambda_i,\lambda_P\rangle$ coming
from the MHV vertices.
Here $\lambda_i$ is the positive chirality spinor of the $i^{th}$ 
external particle
and $\lambda_{P\,a}=P_{a\dot a}\eta^{\dot a}$ is the spinor associated 
with an off-shell
internal particle.  These unphysical singularities are located at 
non-Lorentz invariant points
in momentum space (namely $\lambda_i^aP_{a\dot a}\eta^{\dot a}=0$), so 
they cancel
by virtue of the proof of Lorentz invariance of the sum of MHV tree 
diagrams given in section 5
of \CachazoKJ.  Indeed, diagrams containing unphysical 
singularities appear in
pairs, and they cancel by the pairwise cancellation that was used in 
\CachazoKJ\ to prove
eqn. (5.17) of that paper as part of the proof of Lorentz invariance.

What does the comparison to the BCF recursion relations really add to 
this discussion?
As we have already noted, it was shown in \CachazoKJ\ that
the MHV tree diagrams generate the same singularities as the Feynman 
diagrams.
(Cancellation of unphysical singularities, though it follows from 
Lorentz invariance, was not
stated explicitly.)  Tree amplitudes are completely determined by the 
singularities they
possess when analytically continued to complex momenta.  This statement 
can be
proved by the following reasoning. Tree amplitudes are rational 
functions of the spinor variables $\lambda$ and $\tilde\lambda$.
A rational function of complex variables that has no singularities is a 
polynomial.  Hence,
letting $A$ denote  a tree amplitude computed from Feynman diagrams and 
$\tilde A$
the corresponding amplitude from MHV tree diagrams, if $\tilde A$ has 
the correct
singularities then $A-\tilde A$ is a polynomial.  But on dimensional 
analysis, no such polynomial
is possible for Yang-Mills tree amplitudes with $n>4$ gluons (the 
dimension of an $n$ gluon tree amplitude is $4-n$).  For $n= 4$ gluons, 
the validity of the MHV tree diagrams can be
checked directly.

So the equality $A=\tilde A$ essentially follows from the analysis of 
singularities and Lorentz
invariance in \CachazoKJ.  This approach has one drawback:
 while  a tree amplitude is determined by
its singularities, there has been until now in Yang-Mills theory no 
standard, convenient way to actually
use the singularities to determine the amplitude.  What the BCF 
recursion relations give
us is an extremely convenient way to determine a Yang-Mills tree 
amplitude from its
singularities, making far more satisfying the proof of validity of the 
MHV recipe based on
knowledge of the singularities.

\bigskip

\centerline{\bf Acknowledgments}

Work of R. Britto, B. Feng and E. Witten was supported by NSF
grant PHY-0070928; and that of F. Cachazo was supported in part by
the Martin A. and Helen Chooljian Membership at the Institute for
Advanced Study and by DOE grant DE-FG02-90ER40542. Opinions and
conclusions expressed here are those of the authors and do not
necessarily reflect the views of funding agencies.

\listrefs

\end